\documentclass[10pt]{article}
\usepackage{graphicx}

\def\Title#1{\begin{center} {\Large #1 } \end{center}}
\def\Author#1{\begin{center}{ \sc #1} \end{center}}
\def\Address#1{\begin{center}{ \it #1} \end{center}}

\newcommand\pubblock{\rightline{\begin{tabular}{l} Proceedings of the Second Annual LHCP\\ \pubnumber\\
         \pubdate  \end{tabular}}}

\newenvironment{Abstract}{\begin{quotation} \begin{center} 
             \large ABSTRACT \end{center}\bigskip 
      \begin{center}\begin{large}}{\end{large}\end{center} \end{quotation}}

\newenvironment{Presented}{\begin{quotation} \begin{center} 
             PRESENTED AT\end{center}\bigskip 
      \begin{center}\begin{large}}{\end{large}\end{center} \end{quotation}}

\def\beq{\begin{equation}}
\def\eeq#1{\label{#1}\end{equation}}
\def\eeqn{\end{equation}}

\def\beqa{\begin{eqnarray}}
\def\eeqa#1{\label{#1}\end{eqnarray}}
\def\eeqan{\end{eqnarray}}

\let\bar=\overbar

\def\Dslash{\not{\hbox{\kern-4pt $D$}}}
\def\dslash{\not{\hbox{\kern-2pt $\del$}}}

\def\msb{{\bar{\ssstyle M \kern -1pt S}}}

\usepackage{color}
\usepackage{amsmath}
\usepackage{lineno}

\newcommand\pubnumber{ ATL-PHYS-PROC-2014-110 }

\newcommand\pubdate{\today}

\def\affiliation{
On behalf of the ATLAS Collaboration, \\
Max-Planck-Institut f\"ur Physik \\
F\"ohringer Ring 6, Munich, Germany }

\begin{document}

\large
\begin{titlepage}
\pubblock

\vfill
\Title{  Muon Reconstruction Efficiency, Momentum Scale
and Resolution in $pp$ Collisions at 8 TeV with ATLAS  }
\vfill

\Author{ Maximilian Goblirsch-Kolb  }
\Address{\affiliation}
\vfill
\begin{Abstract}
The ATLAS experiment identifies and reconstructs muons with two high precision tracking systems, the Inner Detector and the
Muon Spectrometer, which provide independent measurements of the muon momentum. This paper summarizes the performance
of the combined muon reconstruction in terms of reconstruction efficiency, momentum scale and resolution. Data-driven techniques
are used to derive corrections to be applied to the simulation in order to reproduce the reconstruction efficiency, momentum scale
and resolution observed in experimental data, and to assess systematic uncertainties on these quantities.
The dataset analysed corresponds to an integrated luminosity of $20.4$~$\text{fb}^{\text{-1}}$ from $pp$~collisions at $\sqrt{s}$ = 8 TeV recorded in 2012.

\end{Abstract}
\vfill

\begin{Presented}
The Second Annual Conference\\
 on Large Hadron Collider Physics \\
Columbia University, New York, U.S.A \\ 
June 2-7, 2014
\end{Presented}
\vfill
\end{titlepage}
\def\thefootnote{\fnsymbol{footnote}}
\setcounter{footnote}{0}
%

\normalsize 


\section{Introduction}

The ATLAS experiment~\cite{{Aad:2008zzm}} is a multipurpose detector at the Large Hadron Collider (LHC). 
One important capability of the detector is the precise identification and reconstruction of muons. 
This is achieved by combining information collected in two subdetectors - the Inner Detector~\cite{Aad:2010bx} (ID) and the Muon Spectrometer~\cite{Aad:2010ag} (MS). Both are capable of measuring charged particle tracks and their momenta using magnetic fields. By combining the measurements, a reliable muon reconstruction is possible in a momentum range reaching from several GeV to the TeV scale. 
In this paper, the muon reconstruction in ATLAS is briefly described. 
Measurements of the reconstruction efficiency and the momentum resolution carried out using collision data~\cite{Aad:2014rra} are presented.
By running the same measurements on simulated collision events, corrections can be derived for calibrating the simulated datasets used in physics analyses.

\section{Muon Reconstruction with the ATLAS Detector}

The ATLAS muon reconstruction makes use of two main detector subsystems:
the ID is the subdetector closest to the beam pipe. It consists of silicon pixel and strip detectors as well as a Transition Radiation Tracker (TRT). A 2 Tesla solenoid magnetic field allows the measurement of charged particle momenta. Apart from track reconstruction, the Inner Detector is also responsible for the identification of vertex candidates. The covered pseudorapidity range of the silicon systems is $|\eta| < 2.5$, while the TRT reach is $|\eta| < 2.0$. 

The MS is the outermost layer of the detector. Situated outside the calorimeters, it is embedded in a toroidal magnetic field of 0.5$\,$T on average. 
Two types of tracking chambers are used:
fast trigger chambers are responsible for assigning the presence of a muon to a specific bunch crossing. They also provide a coarse measurement of the $(\eta,\phi)$ coordinates of passing muons. In the barrel region of the detector ($|\eta| < 1.05$), this task is fulfilled by Resistive Plate Chambers (RPC), while in the endcaps ($1.05 < |\eta| < 2.4)$, Thin Gap Chambers (TGC) are used. 
The second type of chamber are the precision chambers. In most of the detector ($|\eta| < 2.0$), gas-filled Monitored Drift Tubes (MDT) are used as precision chambers. In the forward region of $2.0 < |\eta| < 2.7$, which is affected by a higher background, Cathode Strip Chambers (CSC) are used instead.
The central region of the detector, $|\eta| < 0.1$, is only partially instrumented due to  an access for cabling. 

Several complementary algorithms are available for the identification and reconstruction of muons. 
\begin{itemize}
    \item Combined (CB) muons are a combination of independently reconstructed tracks in the Inner Detector and Muon Spectrometer. Using a statistical combination or a global track refit, a combined track is formed. Combined muons can be reconstructed in detector regions covered by both the ID and the MS, which is the case in $|\eta| < 2.5$.
    \item Segment-Tagged (ST) muons consist of a track in the inner detector that is associated to at least one track segment in a precision chamber layer of the Muon Spectrometer. This muon type can enhance the acceptance of low momentum muons, which may not traverse the entire muon spectrometer.
    \item Standalone (SA) muons consist of a track in the Muon Spectrometer without an associated Inner Detector Track. The track is then extrapolated from the Spectrometer to the interaction point. This muon type can provide acceptance in the forward region of $|\eta| > 2.5$, where the inner detector does not provide a track measurement. 
    \item Calorimeter-tagged (CT) muons consist of a track in the inner detector associated to an energy deposit in the calorimeters compatible with a minimal ionizing particle. These muons are used to improve acceptance in the region of $|\eta|<0.1$, where the Muon Spectrometer is not fully instrumented. 
\end{itemize}

\section{Measurement of the reconstruction efficiency}

The muon reconstruction efficiency describes the probability that a certain muon is successfully reconstructed. 
It is measured using $Z\to\mu\mu$ and $J/\psi \to \mu\mu$ decays, using a Tag-and-Probe method as described in detail in~\cite{Aad:2014rra}.

A comparison of the efficiencies measured in collision data, $\epsilon_{\text{Data}}$, to the result obtained in simulated $Z\to\mu\mu$ or $J/\psi \to \mu\mu$ decays, $\epsilon_{\text{MC}}$, then yields a correction that can be applied to the simulation, the so called 'efficiency scale factor' (SF):
$$ \text{SF} = \frac{\epsilon_{\text{Data}}}{\epsilon_{\text{MC}}}$$

\begin{figure}[htb]
\centering
\includegraphics[width=0.48 \textwidth]{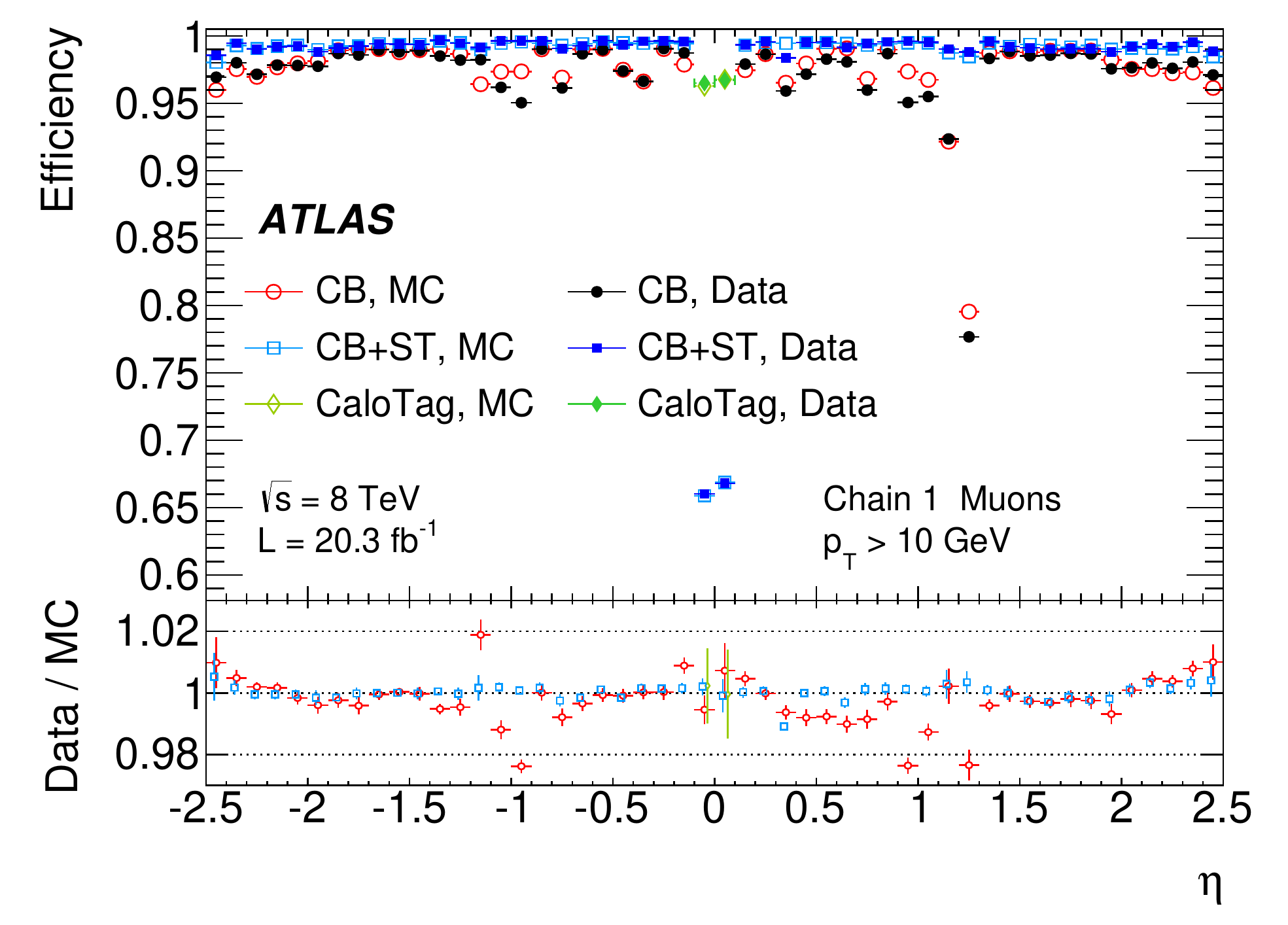}
\includegraphics[width=0.48 \textwidth]{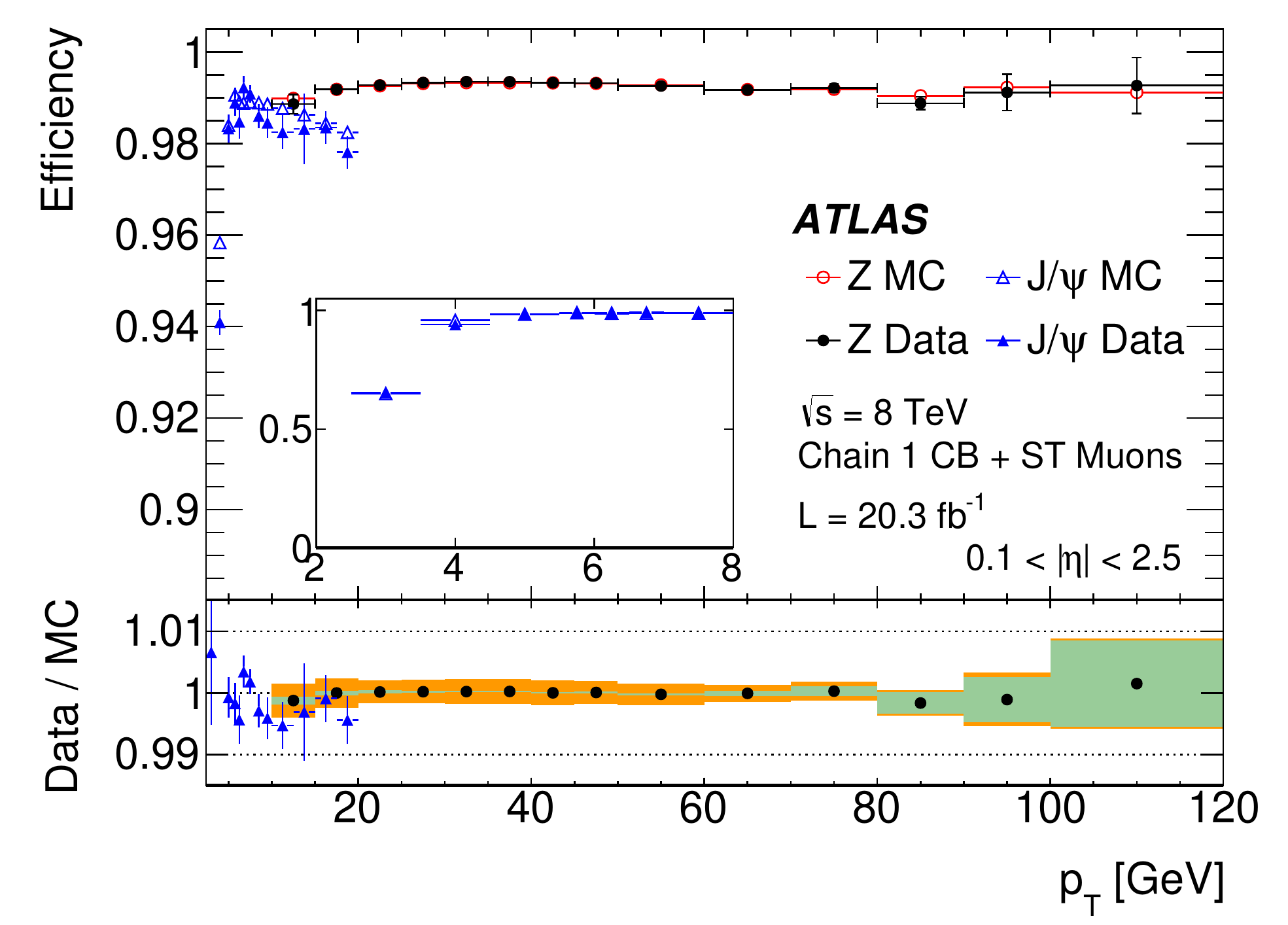}
\caption{Measured muon reconstruction efficiency as a function of the pseudorapidity $\eta$ (left) and the transverse momentum $p_{T}$ (right)~\cite{Aad:2014rra}.}
\label{fig:figure2}
\end{figure}

The measured reconstruction efficiencies for the combination of CB+ST muons (Figure \ref{fig:figure2}), which is the most common combination used in ATLAS physics analyses, is above $98\%$ throughout the entire range with the exeption of the central region of $|\eta| < 0.1$, where due to lacking instrumentation of the MS a drop to $65\%$ is observed. If Calorimeter-Tagged muons are used, this efficiency loss can be almost fully compensated. The efficiencies are found to be stable as a function of the transverse momentum. 
An excellent level of agreement between the simulation and collision data is observed, with an agreement at permill-level through most of the detector. 
If only combined (CB) muons are used, disagreements at a $1\%$ level are observed in the region of $\eta \sim 0.5$, caused by the RPC detector conditions during the data-taking, and at up to $2\%$ level in $0.9 < |\eta| < 1.3$, related to imperfections in the simulation of the transition between the barrel and endcap detector  regions.

\section{Calibration of the muon momentum resolution and scale}
The momentum resolution and scale of the ATLAS muon reconstruction are calibrated using $Z\to\mu\mu$ and $J/\psi \to \mu\mu$ decays. 
The muon momentum resolution can be modelled by a quadratic sum of two terms: 
$$ \frac{\sigma(p_{T})}{p_{T}} = a \oplus b \cdot p_{T}$$ 
Including a scale correction leads to the following form for the calibration correction applied to muons in the simulation:
$$ p_{T}^{\text{corr}} = p_{T}^{\text{MC}} \cdot s \cdot \left( 1 + \Delta p_{1} G(0,1) + p_{T} \cdot \Delta p_{2} G(0,1) \right),  $$
where $s$ is a scale correction and the $\Delta p_{i}$ are momentum resolution parameters. 
Each $G(0,1)$ represents a normally distributed random number with a mean 0 and a width 1. The scale and momentum resolution parameters are determined in 16 different $\eta$ regions of the detector using a maximum likelihood template fit to $Z\to\mu\mu$ events in collision data and simulation.

\begin{figure}[htb]
\centering
\includegraphics[width=0.4 \textwidth]{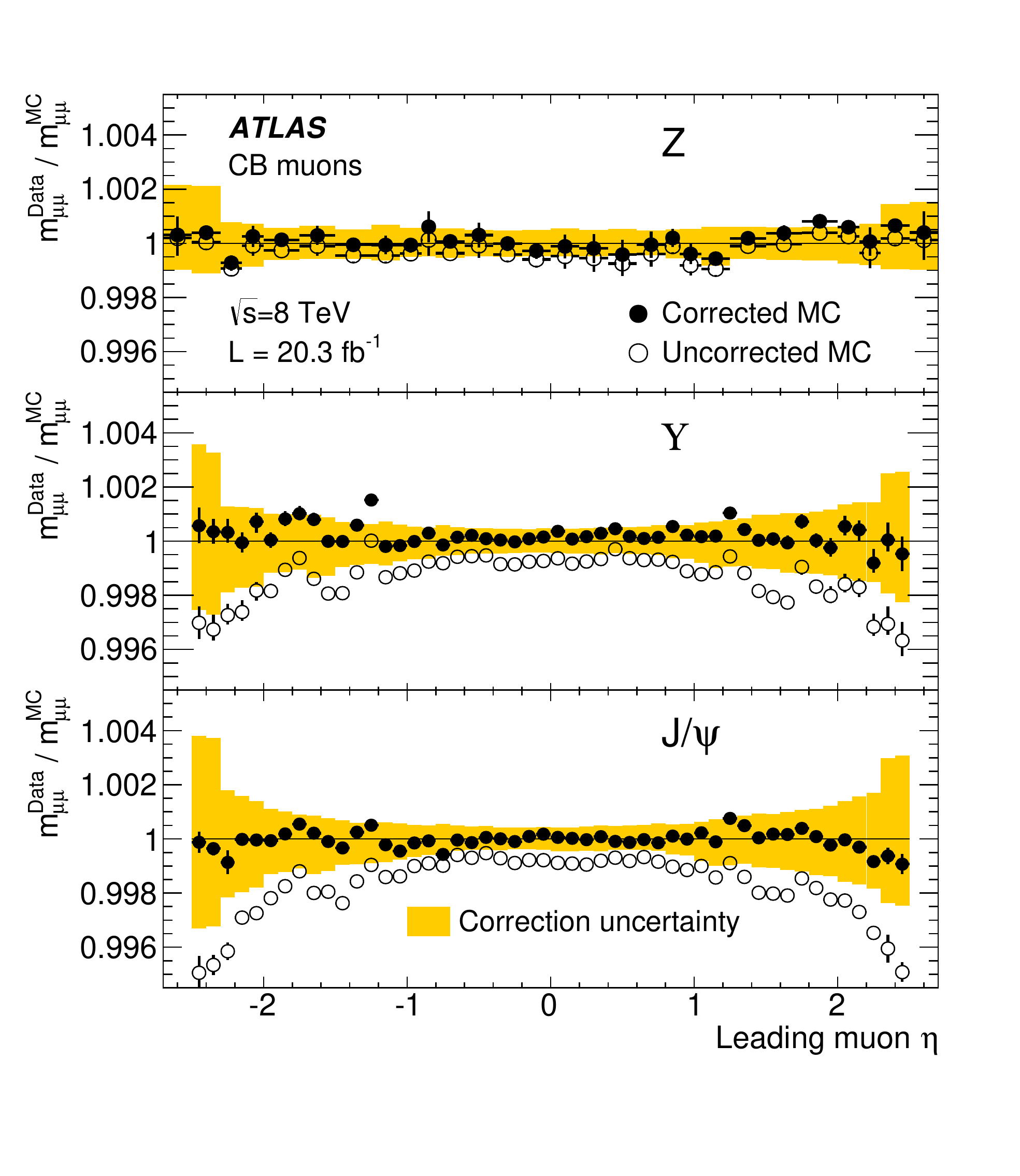}
\includegraphics[width=0.44 \textwidth]{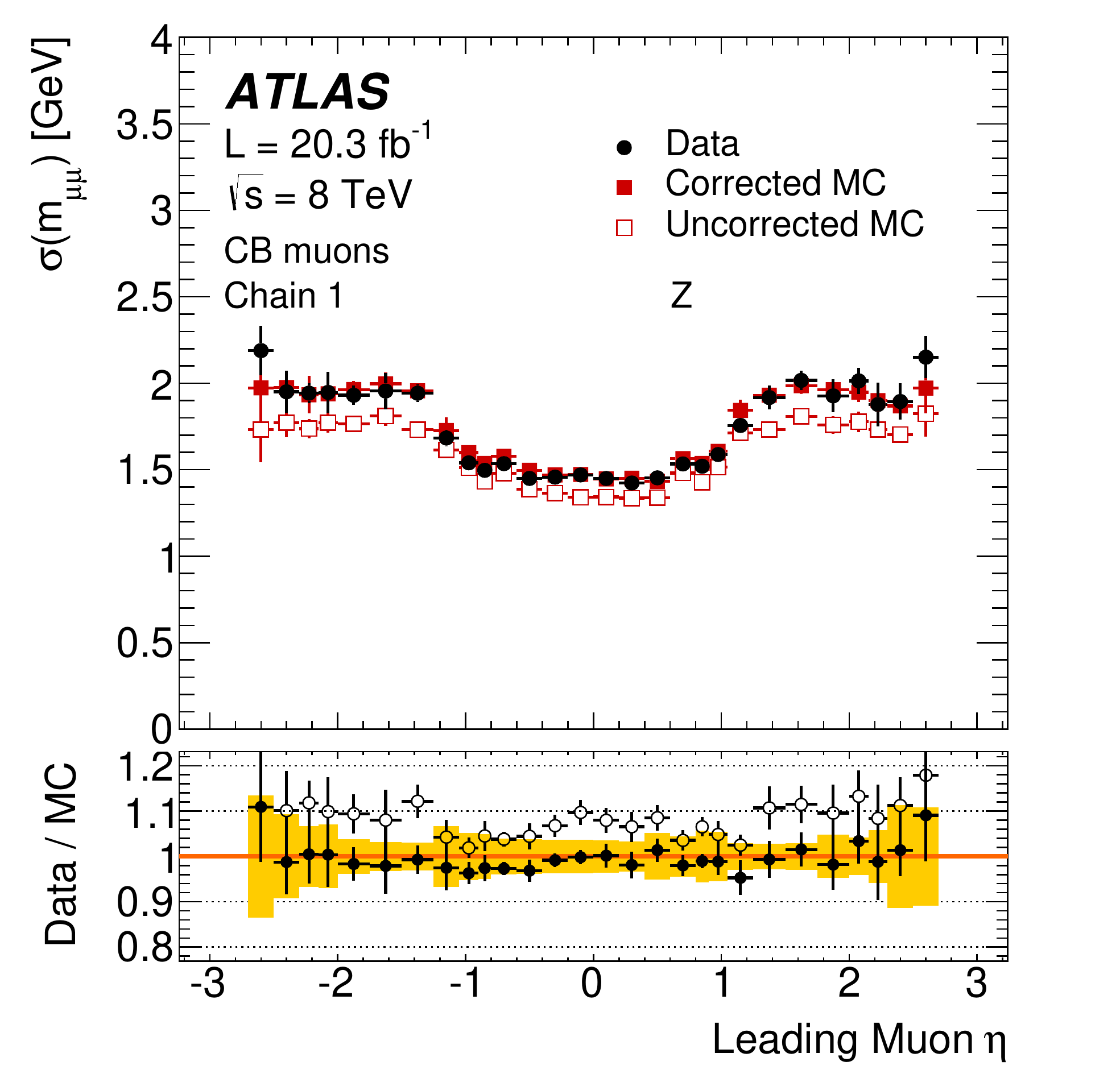}
\caption{Ratio of the reconstructed mean mass for the $J/\psi$,$\Upsilon$ and Z resonances~(left) and invariant mass resolution for the Z resonance~(right) in data and the corrected simulation as a function of the muon pseudorapidity $\eta$~\cite{Aad:2014rra}.}
\label{fig:figure3}
\end{figure}

The resulting calibration is then validated using $J/\psi\to\mu\mu$,$\Upsilon\to\mu\mu$ and $Z\to\mu\mu$ events.
At the Z mass, a momentum resolution (defined as the width of a Gaussian convoluted with the generator-level lineshape) of 1.5-3 GeV in various regions of the detector is obtained. After corrections, the muon momentum scale and invariant mass resolution are found to be in agreement between the data and the simulation, as depicted in Figure \ref{fig:figure3}.

\section{Conclusions}
The ATLAS detector is able to reconstruct and identify muons in the Inner Detector and Muon Spectrometer. Several algorithms are available to maximize the muon acceptance. The most commonly used set of algorithms provides a high and stable efficiency above $98\%$, well described by the simulation. The momentum resolution and scale are also predicted with solid accuracy and require only small corrections.

\end{document}